\documentclass[11pt]{article}
\usepackage{amsmath}
\usepackage{times}
\usepackage{amsfonts}
\usepackage{graphics,graphicx}
\usepackage{subfigure}

\author{Francis Iannacci\\
\texttt{iannacci@cs.washington.edu} \\
Department of  Computer Science and Engineering \\
Seattle, WA, 98195 
  \and 
Yanping Huang\\
\texttt{huangyp@cs.washington.edu} \\
Department of  Computer Science and Engineering \\
Seattle, WA, 98195 
}
\title{ChirpCast: Data Transmission via Audio}

\begin{document}

\maketitle

\begin{abstract}
  In this paper we present ChirpCast, a system for broadcasting network access keys to laptops ultrasonically. This work explores several modulation techniques for sending and receiving data using sound waves through commodity speakers and built-in laptop microphones. Requiring only that laptop users run a small application, the system successfully provides robust room-specific broadcasting at data rates of 200 bits/second.
\end{abstract}

\section{Introduction}

Providing selective access to public wireless networks is an open challenge in computer networks. In many instances, such as at an office or a coffee shop, network access should be granted based on location: those within the physical space should allowed network access, whereas those outside should be denied. At its core, this problem balances the desire of access point providers to limit access, thereby reducing necessary bandwidth and cost, with the desire of users to gain access to networks easily and automatically. At present, solutions to this problem often favor one party at the expense of the other.

It is common practice for businesses to provide free and open wireless network access. While this provides patrons with convenient network access, it does not restrict network access to those in the physical space.  Standard wireless transmission can easily pass through walls and other obstructions, allowing devices outside of the intended space to gain network access. As a result, the access point provider may require more bandwidth and incur its associated higher cost, and the patrons may have their connections slowed by access point freeloading. Thus, to reduce provider costs and increase the connection speed of permitted users, some form of access control is necessary.

Currently, the predominant method for access control~\cite{sandhu1994access} is to require users first to obtain a network access key before being granted network access.  In an office or coffee shop setting, this key is obtained from an employee of the establishment, with network access occasionally contingent on a product purchase. While this is effective is restricting network access to customers and staff, it can be inconvenient for users to be forced into a purchase. Additionally, it does not prevent users from obtaining the passkey and then accessing the network from outside of the designated wireless access space. A passkey distribution system~\cite{frykholm2001error} therefore should have the following properties: (1) it must appear nearly automatic from the user's perspective to be convenient, and (2) it should allow passkeys to be changed sufficiently often to ensure users stay within the designated access zone.

In this paper we present ChirpCast, a physical layer which distributes access keys using ultrasonic transmissions~\cite{weng1993ultrasonic}.  Using inexpensive computer speakers and a small encoding program, the ChirpCast transmitter broadcasts access keys which are inaudible to humans. Unlike radio waves, sound transmissions do not pass through walls, enabling room-level access locality. \cite{Ambrose03} On the receiver side, ChirpCast leverages a laptop's built-in microphone to capture the signal and then process it in software to recover the data. This system operates in real time, allowing the access key to be changed frequently.

Previous research has explored using audio to transmit data between computing devices in many contexts. Modems are an early example of using sound for point-to-point data transmission.  Recently, researchers have explored using audio transmission in context aware computing applications: Madhavapeddy et al\cite{Madhavapeddy03} describes several modulation techniques for audio networking, including a physical layer that uses inaudible sound to transmit data transmission. This research demonstrated a data rate of 8 bits/s with 95\% accuracy.   Our project is an extension of this work, examining new modulation techniques for more noise-immune and faster transmissions.

The structure the remainder of the paper as follows: Section \ref{sec_carrier} describes the selection of carrier frequency, Section \ref{sec_modulation} discusses our modulation techniques and results, and Section \ref{sec_conclusion} describes our findings and provides directions for future work.

\section{Carrier Frequency Selection}
\label{sec_carrier}
The use of sound as data carrier provides an excellent means of room-specific broadcast localization, since sound is attenuated by physical barriers. \cite{Ambrose03} Beyond localization, we need our carrier to have the properties that (1) it is inaudible to adults; and (2) the speakers and microphones are capable of delivering and receiving high power at the chosen discrete frequencies.

We conducted a small study to find the limits of adult hearing~\cite{blamey1992factors}, whic consisted of four adult participants, three males and one female.  Literature often cites 20 Hz to 20kHz as the audible frequency range for humans.  \cite{Ambrose03}  However, the audible range for adults is often less than this, a fact exploited by certain MP3 encoding formats. \cite{Yoon06}  Our study found that no participant could detect frequencies above 17.75 kHz.  We therefore choose carrier frequencies above 18 kHz to ensure they cannot be heard.
	
Next, we characterized the speaker-microphone pair's performance across the sound spectrum.  Our sound card supports transmission and sampling at 44.1 kHz, which according to the Nyquist-Shannon sampling theorem allows a maximum frequency of 22.1 kHz frequency to be sent on and recovered from the channel. We employed adaptive kernel filter~\cite{zhao2012fixed, zhao2012adaptive} to process the signals.  We therefore restrict our frequency characterization to the range between 18 kHz and 22.1 kHz.  Broadcasting pure sinusoidal tones through the speaker and measuring the signal amplitude at the receiver indicates that transmissions are differentiable from noise up to 19.5 kHz. This result is similar to that achieved in \cite{Madhavapeddy03}.

\section{Modulation Techniques}
\label{sec_modulation}

\subsection{Frequency Shift Keying}

Frequency Shift Keying (FSK)~\cite{ghovanloo2004wideband} transmits information by changing the frequency of the channel carrier.  For an audio channel $Cn$, the transmitter broadcasts at frequency $F_Cn(0)$ when transmitting a bit value of 0, and at frequency $F_Cn(1)$ when transmitting a bit value of 1.  This scheme offers greater noise immunity than the simpler On-Off Keying, since the absence of both $F_Cn(0)$ and $F_Cn(1)$ during a transmission indicates that an error has occurred.   Additionally, since noise events affect the entire 18-19.5 kHz band (refer to Section \ref{sec_practical}), noise events will be recognized and the corrupted data ignored. In this experiment we choose $F_C0(0)$ to be 18 kHz, $F_C0(1)$ to be 18.25 kHz, $F_C1(0)$ to be 18.5 kHz, and $F_C1(1)$ to be 18.75 kHz.

\begin{figure}[h!]
\centering
\includegraphics[scale=0.35]{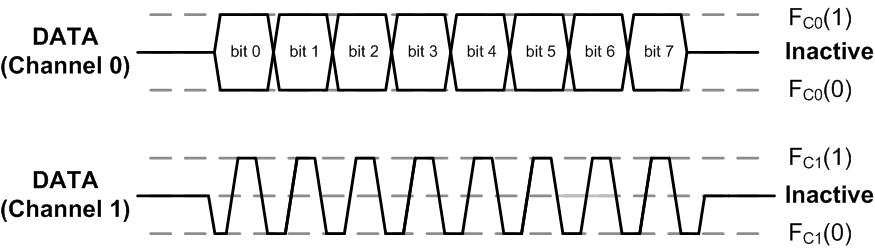}
\label{fig_FSK_Encoding}
\caption{Timing diagram for FSK.  The values on the right side indicate the frequency of the carrier wave.}
\end{figure}

Our implementation consists of two simultaneously transmitted bit streams, a DATA signal on the left speaker channel and a CLOCK signal on the right signal channel. The decision to send a separate clock signal simplifies the sender and receiver synchronization, since the clock signal informs the receiver when it should sample the data without requiring a receiver-maintained reference clock. The timing diagram for this encoding is shown in Figure \ref{fig_FSK_Encoding}. It is worth noting that two streams are the maximum that can be transmitted concurrently, since each signal must be given a separate audio channel to prevent audible aliasing artifacts.  

To recover the data bits from the ultrasonic transmission, the receiver samples the audio from its microphone and stores these samples in a buffer.  Once the buffer is full, the receiver computes the fast Fourier transform (FFT) of the buffer. The average power across each frequency in the range of 18 kHz to 19.5 kHz, excluding carrier frequencies, is computed to find the average noise power per frequency. This is the adaptive noise threshold. The receiver then compares the power at each carrier frequencies against this measure, and if a carrier frequency has power that is an order of magnitude above the adaptive noise threshold then it is identified as being active.

This scheme was successful in transmitting data across a distance of one meter at a rate of 4 bits/second with over 90\% bit accuracy. This is less than the reported performance of \cite{Madhavapeddy03}. This is due in large part to the speed of the code, which can perform at most eight FFT operations per second. As future work, performing carrier frequency-specific power analysis should be investigated for software speedup.

\subsection{Phase-shift keying}
In phase-shift keying (PSK), we use the phase of a carrier wave to convey data. Let $m(t)$ be the network access key we would like to broadcast, $m(t) \in \{\pm 1\}$, where $+1$ indicates a logical one and $-1$ indicates a logical zero. The value of $m(t)$ change every $T_b$ seconds, as shown by the red dashed line in the Figure \ref{fig_BPSK}.

\subsubsection{Binary phase-shift keying} 
\begin{figure}[h!]
\centering
\includegraphics[scale=0.15]{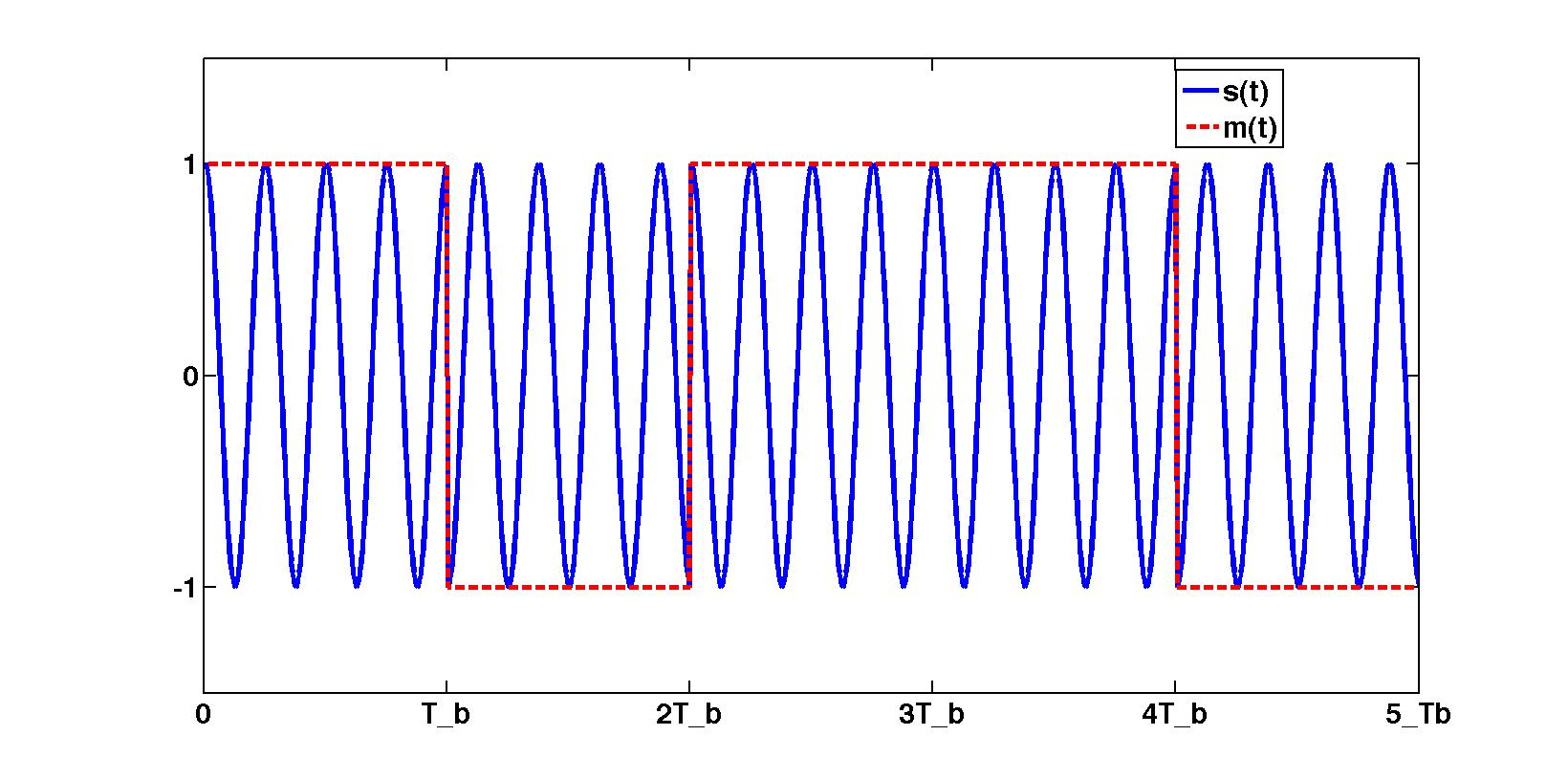}
\label{fig_BPSK}
\caption{BPSK modulated signal s(t) and the data sequence m(t)}
\end{figure}

Our first attempt at modulating $m(t)$ with PSK is to multiply $m(t)$ by a sinusoidal carrier wave at $\omega_c$
\begin{equation}
  \label{eq_BPSK}
  s(t)  = A m(t) \cos(2 \pi \omega_c t)
\end{equation}
where A is the amplitude of the carrier wave. Since $-\cos(2\pi\omega_ct) = \cos(2\pi\omega_ct + \pi)$, the carrier has either a $0$ degree or a $180$ degree phase shift depending on the data, a binary phase shift keyed (BPSK) signal. The solid blue curve in figure \ref{fig_BPSK} shows the modulated signal $s(t)$. 

Demodulation of BPSK signal is simple if the receiver has a clock reference $c(t)$ that is at the exact phase and frequency as the carrier wave of the sender, $c(t) = \cos(2\pi \omega_ct)$. Suppose the receive signal $r(t)$ is the sum of $s(t)$ with some white noise corruption $n(t)$ and there is no propagation delay between the sender and the receiver. The demodulation is carried out by a convolution of $r(t)$ with $c(t)$
\begin{equation}
  \label{eq_BPSK_demo}
  y(t) = \int_o^{T_b} r(t-\tau) c(\tau) d\tau 
\end{equation}
The value of $y(t)$ sampled at the end of each bit period is used to determine the demodulated binary output. $y(nT_B) > 0 $ indicates that the $n$-th symbol is a logical one  while $y(nT_B) < 0$ indicates a logical zero.

In reality, the distance between the speaker and the microphone is not fixed so that the propagation delay cannot be known by the receiver in advance. Therefore, the clock reference is not at the same phase as the carrier waveform. To deal with this problem, we insert an initial header sequence $m_0(t)$ at the beginning of each data sequence $m(t)$. The initial $m_0(t)$ is also known by the receiver. In this way, we could use various supervised learning techniques, such as linear regression, to estimate the unknown propagation delay. However, this approach is subject to the following problems. First, linear regression cannot be performed in real time. The receiver needs to store $r(t)$ with length greater than the length of $m_o(t)$. Second, linear regression is computation expensive since it involves taking the inverses of matrices. Third, the quality of estimation depends on the length of $m_0(t)$. The extra $m_o(t)$ introduces a unnecessary redundancy to the transmitted data. 

\subsubsection{Differential Phase Shift Keying }

\begin{figure}[h!]
\centering
\subfigure[DPSK modulated signal s(t) and the data sequence m(t)]{
\includegraphics[scale=0.12]{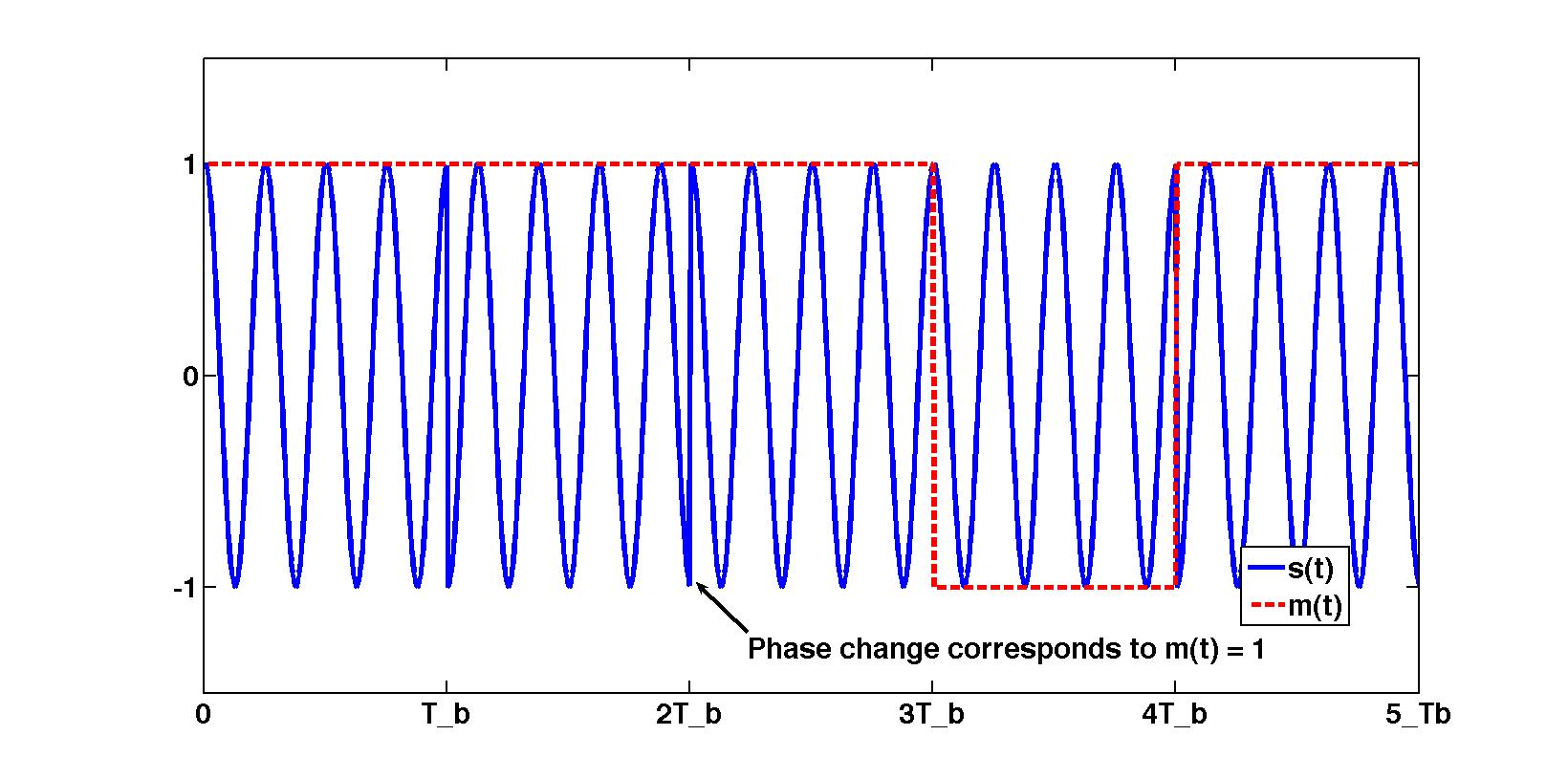}
\label{fig_DPSK}
}
\subfigure[Demodulation of DPSK]{
\includegraphics[scale=0.5]{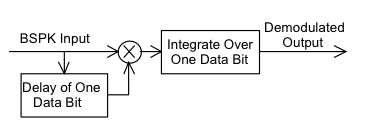}
\label{fig_DPSK_Demo}
}
\caption{DPSK}
\end{figure}

If it is difficult to estimate the propagation error,  it is the similarly difficult to determine whether the demodulated $y(nT_b)$ corresponds to a logical one or a logical zero. However, it is easier to determine if the current estimated phase differs from that of the previous bit. In differential phase-shift keying (DPSK) modulation~, the data signal $m(t)$ is conveyed by changes in the phases of the carrier wave~\cite{rappaport1996wireless}. For example, a logical one may correspond to adding $\pi$ to the current phase and a logical zero may correspond to adding $0$ to the current phase, as shown in figure~\ref{fig_DPSK}.

Demodulation of DPSK signal~\cite{Proakis00} is based on the block diagram of figure~\ref{fig_DPSK_Demo}. The receiver doesn't know the exact propagation delay and the phase of the carrier wave upon receipt. Instead, it simply buffers the received signal every $T_b$ seconds. Suppose the phase difference between $r(t)$ and $r(t-T_b)$  is $\theta$. Then product between current received signal and the signal received one $T_b$ earlier is given by 
\begin{eqnarray}
  \label{eq:DPSK}
  h(t) &\propto& \cos(\omega_c t) \times \cos(\omega_c t + \theta) \nonumber \\
  &=& \frac{1 - \cos(2\omega_c t)}{2} \cos(\theta) - \frac{\sin(2 \omega_c t)}{2}  \sin(\theta) \nonumber\\
\end{eqnarray}
Since the phase change is either $0$ or $\pi$, the $\sin(\theta)$ in the second term can be ignored. Integrating $h(t)$ over $T_b$ seconds gives us the desired binary output $y(nT_b) = \int_{(n-1)T_b}^{nT_b}  h(t) = \cos(\theta)$, where $\cos(0) = 1$  and $\cos(\pi) = -1$.

Using the demodulation method described in equation (\ref{eq:DPSK}), we would be able to recover the original data sequence. The receiver only needs to buffer the received signal for $2T_b$ seconds. The computational complexity is only $O(T)$ for received signal with length $T$.

\subsubsection{Practical considerations}
\label{sec_practical}
\begin{figure}[h!]
\centering
\includegraphics[scale=0.40]{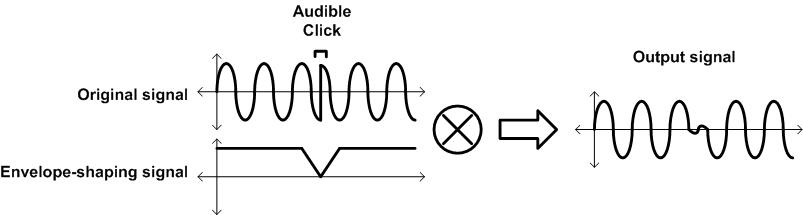}
\label{fig_impulse}
\caption{Technique for removing audible broad-spectrum noise from signal transmission.}
\end{figure}

For the Phase Shift Keying experiments, we choose the carrier frequency to be $\omega_c = 19.2$kHz, which is not audible to most adults. Note that a phase change of $\pi$ in $s(t) $ would lead to a sudden distortion of the waveform, which in turn excites responses over a very broad of frequencies, including those in the audible range. Therefore, it would produce a click-like sound at the end of a bit period when $m(t) = 1$. To remove these artifacts, we reduce the amplitude of $s(t)$ at those phase-alternating times so that the click-like sounds are negligible. This is shown in Figure \ref{fig_impulse}. We only decrease the amplitude at these transition points so that overall signal power is not significantly affected.  Since the amplitude of a signal is independent of the phase, this reduction does not change the modulation or demodulation of the data.

We implement the modulation and demodulation algorithms in Matlab. We use Altec Lansing VS1520 speakers to broadcast the DPSK-modulated signal. The receiver is the build-in microphone equipped in most laptops. The sample rate for both the speaker and the receiver is set at $96$kHz in software.

\subsubsection{Experiments}
\begin{figure}[h!]
\centering
\subfigure[]{
\includegraphics[scale=0.105]{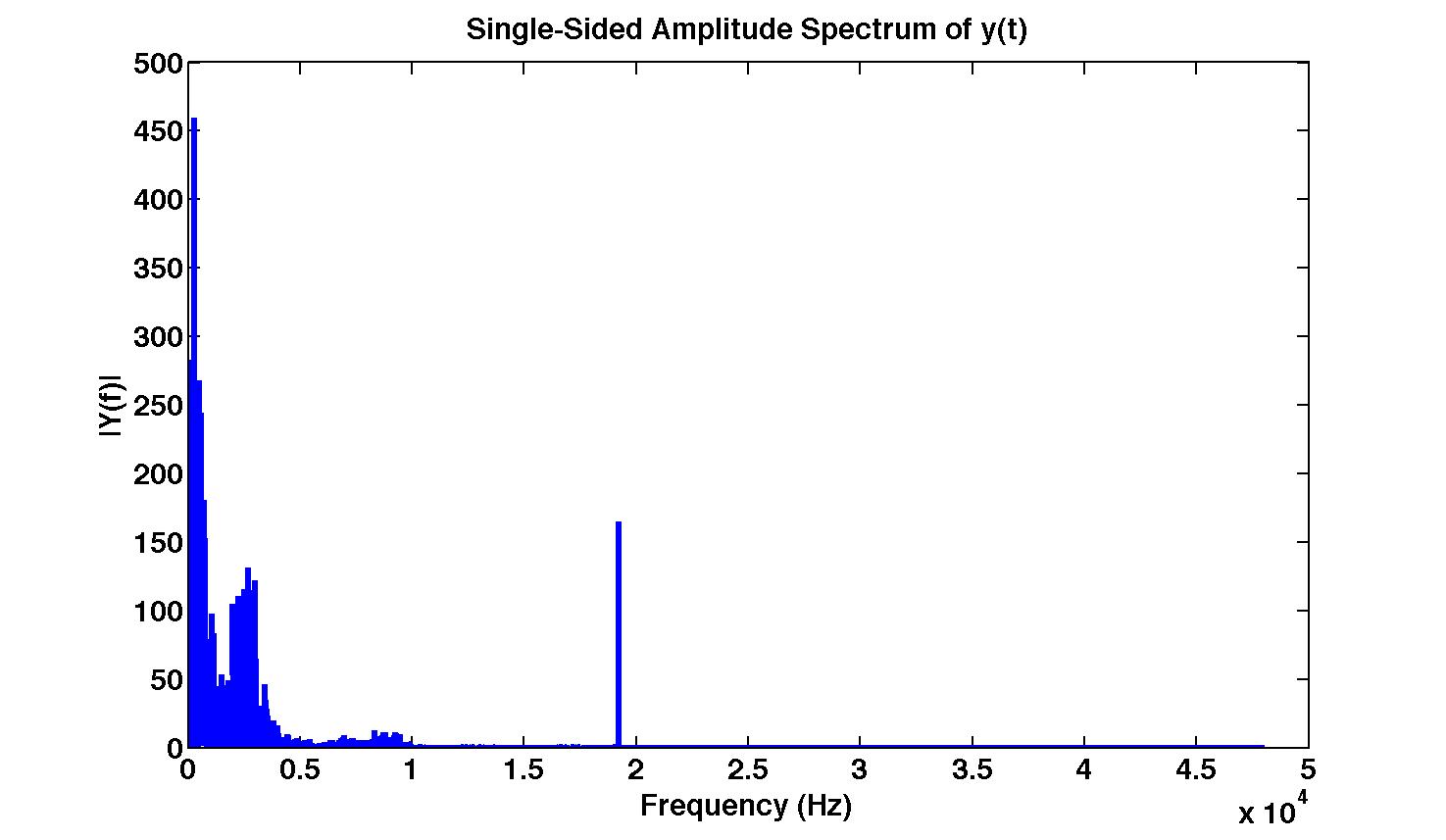}
}
\subfigure[]{
\includegraphics[scale=0.11]{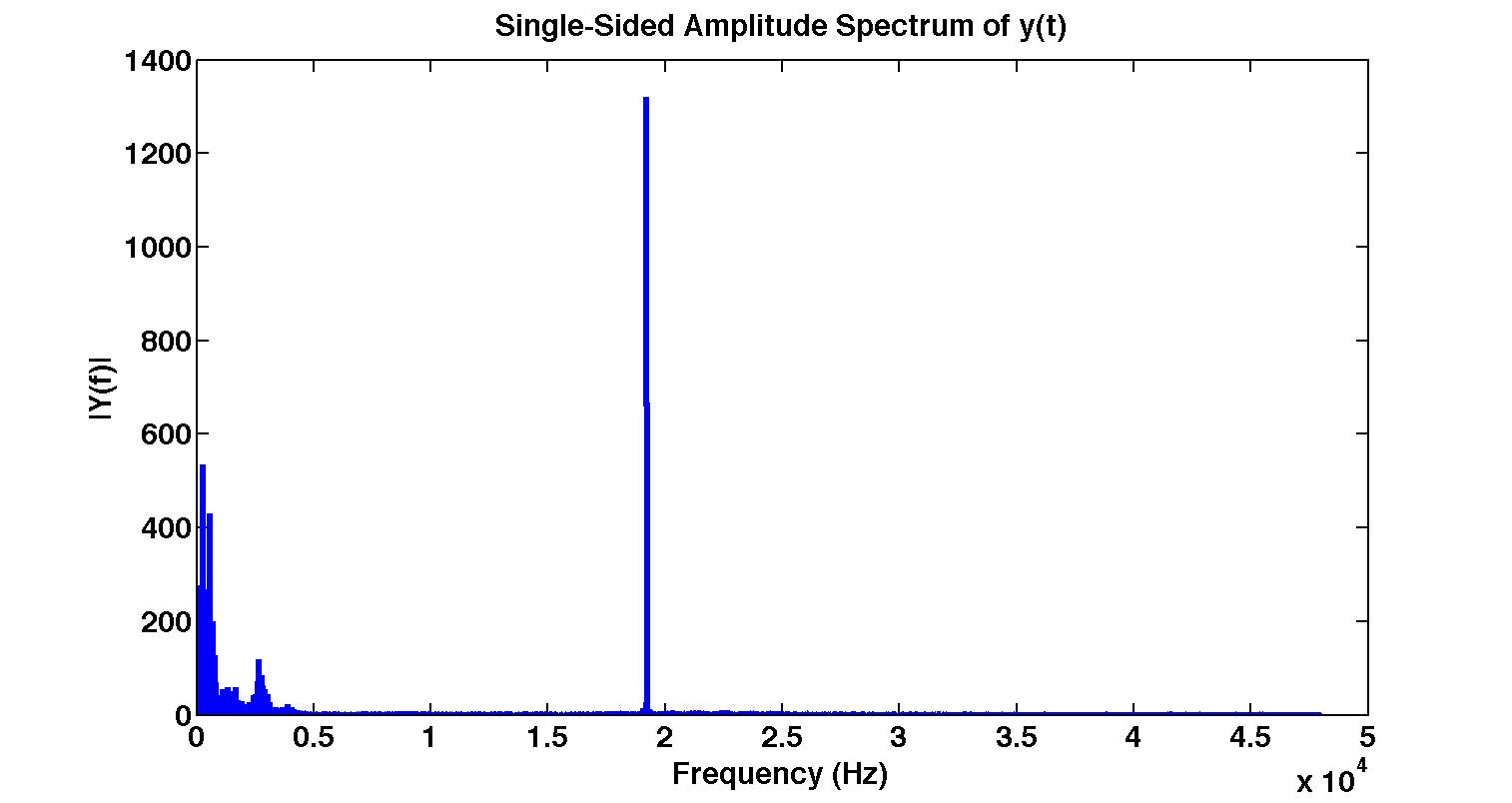}
}
\subfigure[]{
\includegraphics[scale=0.12]{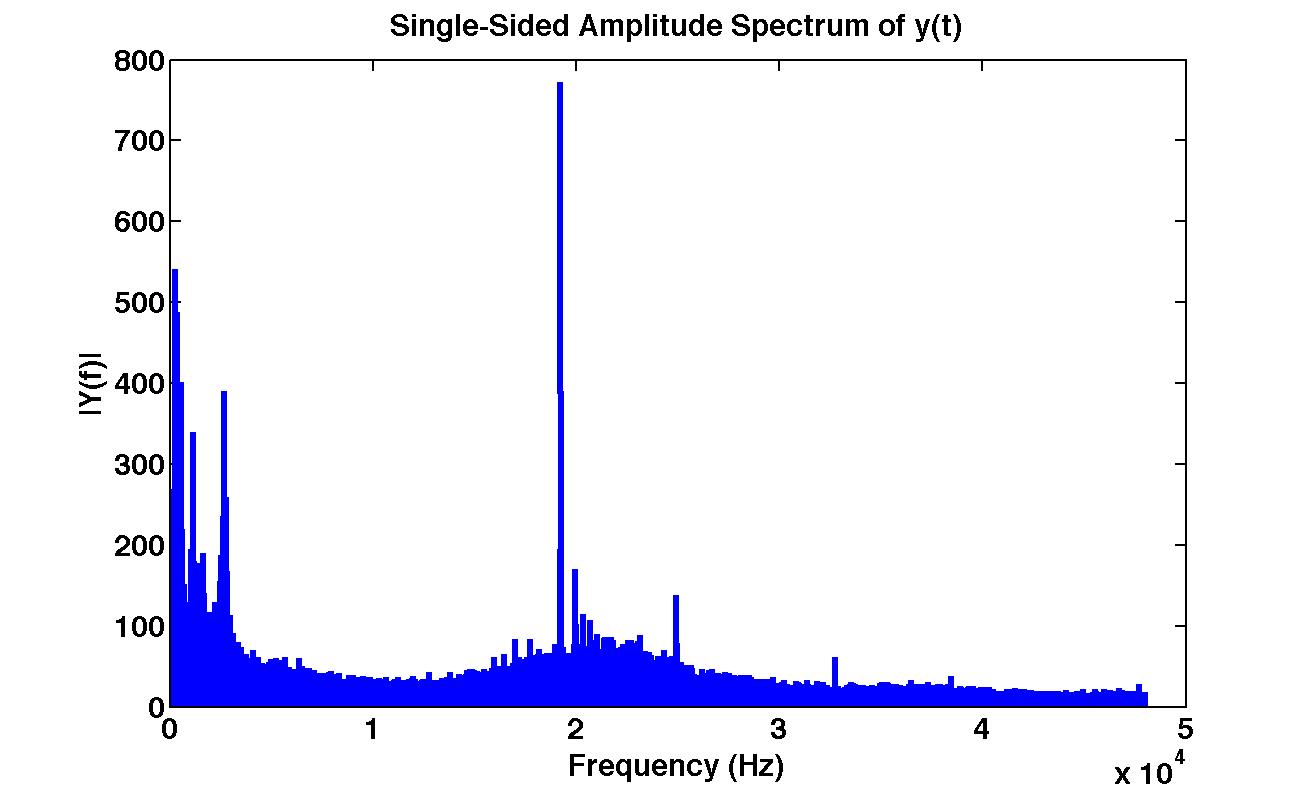}
}
\subfigure[]{
\includegraphics[scale=0.12]{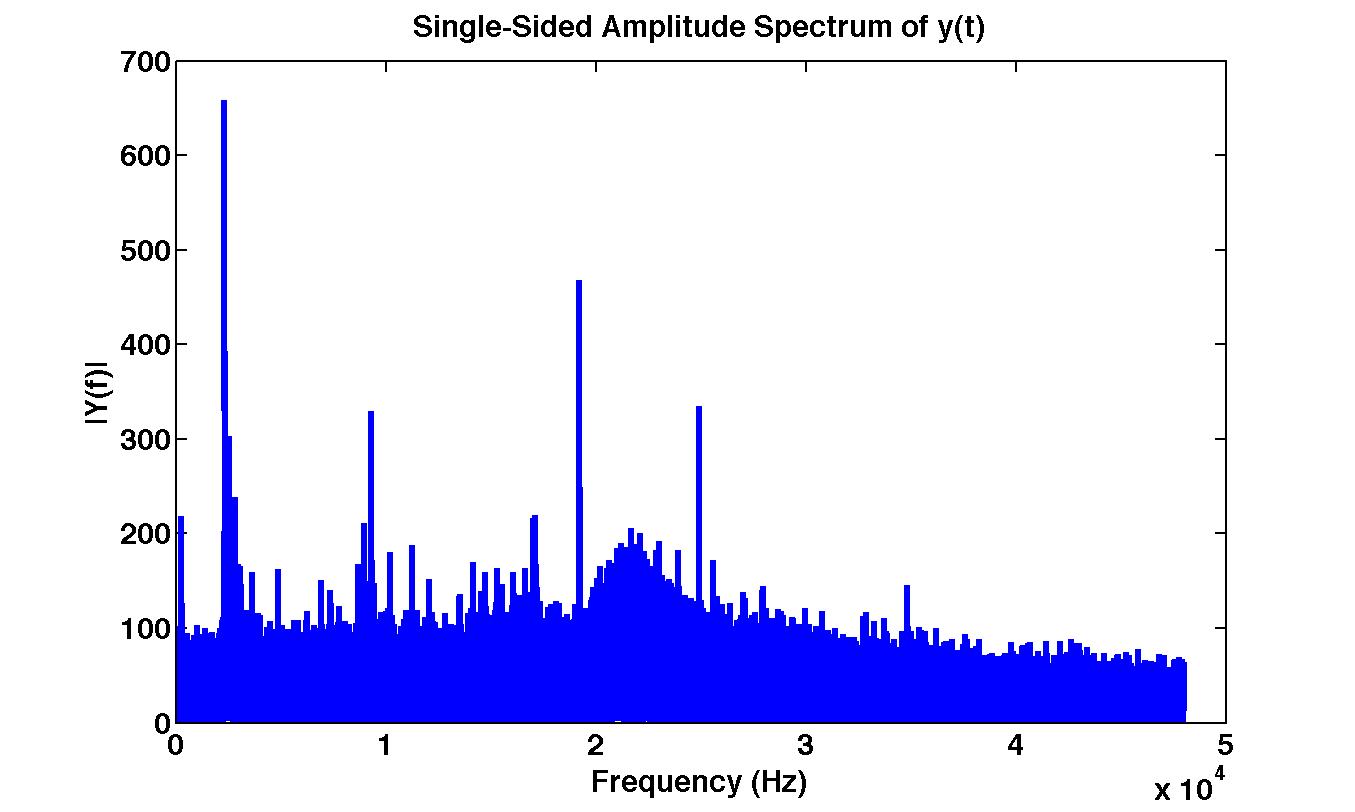}
}
\caption{Power spectrum of received signal $r(t)$ under different background noises. (a) Music. (b) Conversation. (c) Laughter. (d) Key Jangling.  The spike at $19.2$kHz indicates the strength of the data signal. }
\label{fig_enviroments}
\end{figure}

\begin{figure}[h!tb]
\centering
\subfigure[]{
\includegraphics[scale=0.13]{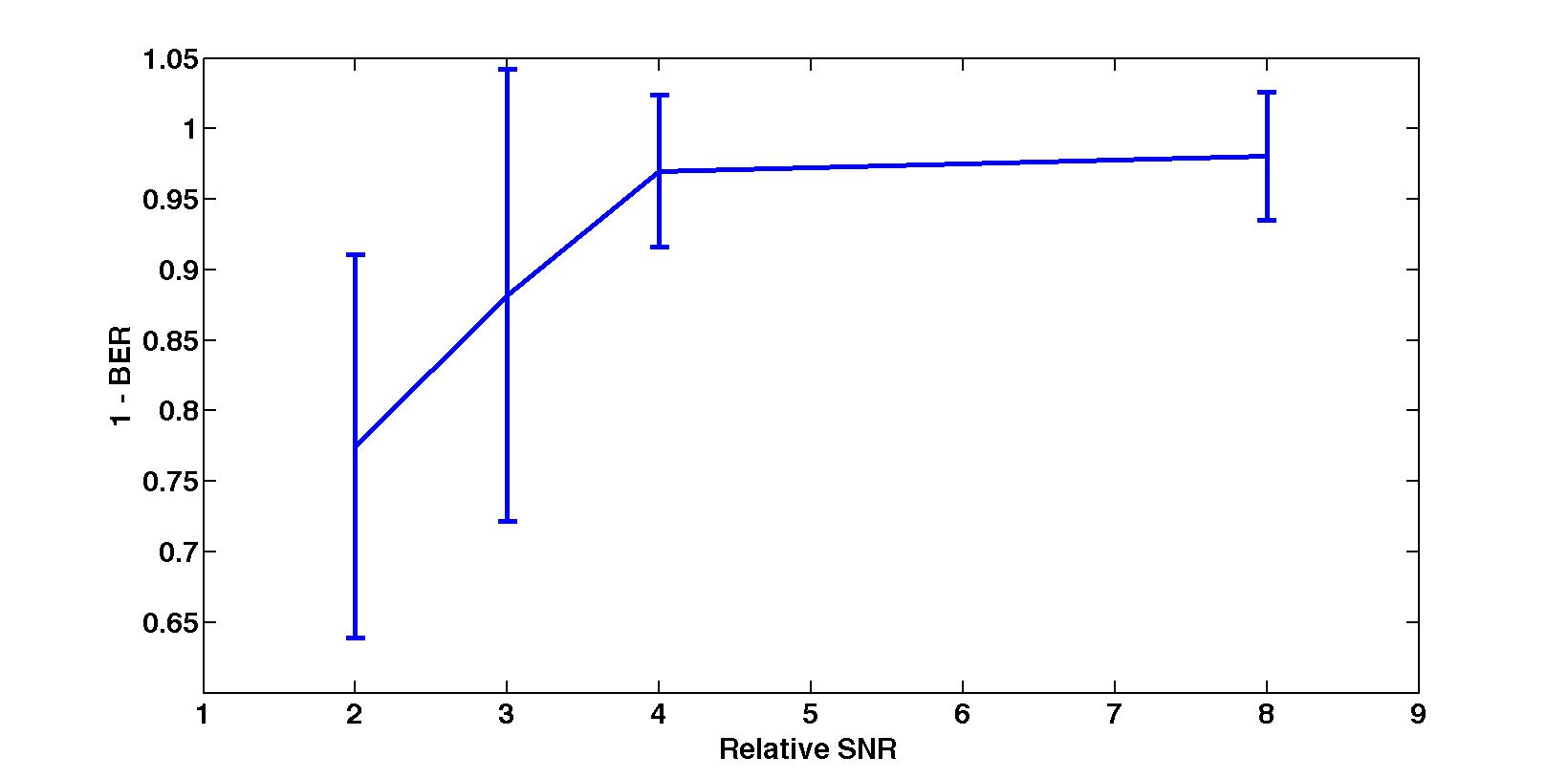}
\label{SNR}
}
\subfigure[]{
\includegraphics[scale=0.13]{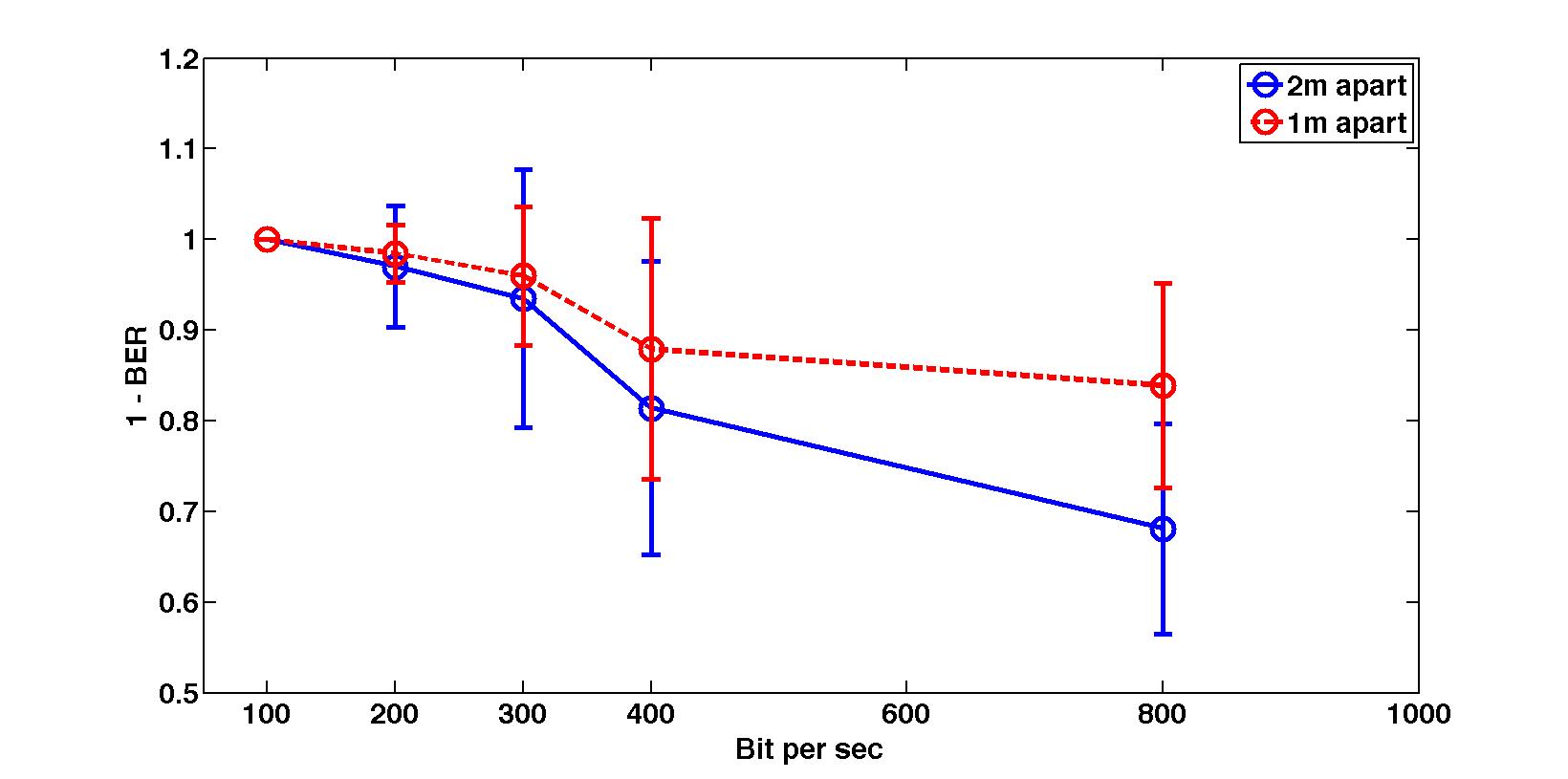}
\label{BPS}
}
\caption{Probability of bit transmission success vs. bit transmission rate when the sender and the receiver are 1m apart (the red dotted line) and 2-meter apart (the blue solid line).}
\end{figure}
Data transmission via DPSK modulated audio signals should be tolerant to background noises in a variety of environments. We tested the robustness of our program under the following typical human activities common in coffee shops or office: playing music, making conversations, uncontrollable laughing, and  key jangling near the microphone. Figure~\ref{fig_enviroments} shows the power spectrum of the received signals under those circumstances during data transmission. Most of the power for music and conversation is below $10$kHz; playing music or making conversation with friends and colleagues won't affect data transmission at $19.2$kHz. Both laughter and key jangling have significant power over a larger range. Laughter's power still mainly resides in the low frequency range, while key jangling's spectrum is flatter, indicating power is more evenly distributed over the frequency range. Thus, the key jangling would have a more damaging effect on the data transmission reliability as it produces relatively more power at the carrier frequency.

The robustness of data transmission is dependent on the signal to noise ratio (SNR) between $s(t)$ and the amplitude of the background noise at $\omega_c$. We compute the bit transmission success rate at different SNR levels and show the results with error bars in figure~\ref{SNR}.  The means and standard deviations are taken over 10 trials. For each trial, we transmit data sequence at $200$ bit per second for $4$ seconds lone, while the speaker and the microphone is placed 1 meter apart. The bit transmission success rate (BTSR) is defined as the ratio between the number of correct demodulated bits and the number of transmitted bits.  \footnote{Note that a single bit error during DPSK demodulation would change the signs of all the following bits.  Therefore, BTSR  shown in figure \ref{SNR} times the total transmitted length can be viewed as the expected length of successful data transmission without any error.  Using geometry distribution, the actually BER is actually $\frac{1}{BTSR \times n}$ where $n = 800$ is the total length of transmitted data in a single trial.} Figure~\ref{SNR} shows an exponential increase in BTSR with respect to SNR. Since modulated signal $s(t)$ is inaudible, we could increase the amplitude of $s(t)$ to achieve high BTSR without affecting most adults (dogs could be in trouble since they have a wider audible range). 

Another factor that affects the performance of audio data transmission is the bit transmission rate. Demodulating of DPSK signal involves an integration over $T_b$ seconds. The larger $T_b$ is, the more accurate the binary output is. At a fixed sampling rate, $T_b$ is inversely proportional to the bit transmission rate. Figure~\ref{BPS} illustrates the relationship between data transmission success rate and the bit transmission rate. We find that ChirpCast can achieve audio data transmission with at least $90\%$ accuracy with the maximum bit transmission rate $=200$bps, when the sender and the receiver are $2m$ apart.

\section{Conclusion}
\label{sec_conclusion}

In this project we have explored different modulation methods to transmit data via audio signals. With at least $90\%$ bit transmission accuracy, we could achieve real time data transmission at a maximum bit transmission rate of $4$bps using frequency-shift keying,  and at a maximum bit transmission rate of $200$bps using differential phase-shift keying. A simple extension to the current project is to use DPSK over multiple orthogonal frequencies simultaneously.  With $k$ frequencies, we could expect a $2^k$ times increase in the maximum bit transmission rate.
\bibliographystyle{unsrt}
\bibliography{sample}
\end{document}